\documentclass[12pt]{article}
\usepackage{latexsym}
\usepackage{amsmath}
\usepackage[dvips]{graphicx}
\usepackage{textcomp}

\newcommand{\be}{\begin{equation}}
\newcommand{\ee}{\end{equation}}
\newcommand{\ba}{\begin{array}}
\newcommand{\ea}{\end{array}}

\author{Fabio Cardone $^{1,2,3}$, Roberto Mignani $^{2,3,4}$,
Andrea Petrucci $^{3,*}$\\ \\ $^{1}$Istituto per lo Studio dei
Materiali Nanostrutturati (ISMN — CNR) \\ Via dei Taurini - 00185
Roma, Italy
\\ $^{2}$GNFM, Istituto Nazionale di Alta Matematica "F.Severi" \\ P.le A.Moro 2 - 00185 Roma, Italy  \\
$^{3}$Dipartimento di Fisica "E.Amaldi", Universit\`a degli Studi
"Roma Tre" \\ Via della Vasca Navale, 84 - 00146 Roma, Italy \\$^{4}$ I.N.F.N., Sezione di Roma III, Italy \\
* Corresponding author: petruccia@fis.uniroma3.it}
\date{}
\title{Remarks on the cavitation of Thorium-228}
\begin{document}
\maketitle \abstract{In this short note we would like to provide
some useful remarks on our previous work about the piezonuclear
decay of Thorium and in general about the methods and protocols that
we used in the experiments on piezonuclear reactions. The purpose of
these remarks is to highlight the critical points of the experiments
and equipment in order to design future experiments that may obtain
positive evidences or that can be as more comparable to previous
ones as possible.}

\section{Highlight of critical points}
It is useful to provide some further remarks and clarifications
about piezonuclear reactions and above all about the experimental
protocols that were implemented in our experiments. Due to the
initial level of studying and understanding of piezonuclear
reactions, the purpose of these notes is to highlight the critical
points of the protocols in order to design future experiments that
would be as much comparable as possible to those experimental
attempts that produced positive results. Let us refer to the papers
that deal with the cavitation of solutions of
Thorium~\cite{thor1,thor2}. In these papers we reported a remarkable
reduction of the initial concentration of Thorium in solutions
treated by ultrasounds and cavitation. In particular the Thorium
content halved in 90 minutes. We have noticed that a few times there
has been a misunderstanding of this statement. In particular, the
following implication was wrongly made: "\emph{\textbf{if}} the
content of Thorium halved in 90 minutes \emph{\textbf{then}} there
must have been an increase of its activity". We have never stated
something like this\footnote{Although the unfortunate first title of
the paper "Speeding up Thorium decay" (as it still is for the Arxiv
version) was misleading. This title was then changed into the more
precise "Piezonuclear decay of Thorium"~\cite{thor1}} and
in~\cite{thor2} we tried to clarify it. According to our
theory~\cite{carmin1,carmin2}, piezonuclear reactions affect nuclei
and induce reactions (nucleo-lysis and nucleo-synthesis\footnote{We
do not use words like \emph{fission}, \emph{fusion}, \emph{decay}
and not even \emph{nuclear reactions} since they all bring with them
concepts which do not suit the anomalous collected evidences.}) of
both stable nuclides~\cite{piezoneutr,neutrpiezo,eur,cavitwater} and
radio nuclides~\cite{thor1} by different mechanisms from those
involved in the well known nuclear processes. In particular in all
of the experiments, where ultrasounds and cavitation were applied to
stable nuclides like Iron~\cite{piezoneutr,neutrpiezo}, the
evidences showed neutron emission from Iron without the usual
following emission of gamma radiation. In this sense the evidences
reported in~\cite{thor1} must be read just like a
\emph{transformation} of Thorium, induced by cavitation, into
something else \footnote{which has not been identified yet} and not
a faster Thorium decay through weak interaction due to an increase
of its activity\footnote{Of course, in the solution the Thorium
nuclei not directly involved in cavitation do continue to decay via
weak interaction, while those involved by piezonuclear reactions
undergo a transformation that without radiation turns them into
other nuclides, maybe stable ones.}. This is why we focused our
measurements on the concentration of Thorium and performed it both
by a mass-spectrometer and by CR39 according to the procedure
described in~\cite{thor3}. Both of these techniques agreed on the
halving of Thorium concentration. In this sense and according to how
piezonuclear reactions are expected to act on nuclei\footnote{which
is something only partially predicted by the theory but whose
behaviour can  be certainly extrapolated and induced by
experiments~\cite{piezoneutr,neutrpiezo,eur,cavitwater}}, it is
wrong to expect an increase of the activity of the solution of
Thorium (or of any other radionuclide) either measured by alpha or
beta or gamma radiation during cavitation. Conversely, on the basis
of the results obtained so far, one is induced to expect a decrease
of the activity of the solution since the involved radionuclide gets
\emph{transformed} by piezonuclear reactions (induced by cavitation)
into something else (that, according to some hints of our theory,
should be stable nuclides). However, in order not to perform
experiments being biased by expectations that might be wrong or
misleading, one has to leave open both chances\footnote{More
mechanisms might be concurrent which might bring about different
transformations or decays of the
radionuclide~\cite{thor1,Filippov1,FilipUrut} with or without
emission of radiation.}. Let us concentrate now on the experimental
equipment, that have been used so far in our experiments. At a first
sight the equipment appears to be very simple: a cavitator which
might sound like a synonym of ultrasound machine; a reaction
chamber, or cavitation chamber which might sound like one of those
normal vessels used in chemistry. The cavitator is indeed an
ultrasound machine and the reaction chamber is indeed a normal
vessel of those ones used in chemistry. However, this circumstance
does not mean at all that these experiments can be replicated by any
ultrasound machine, capable of the same frequency and power, and a
glass bottle or beaker. As we will try to make it clear in this note
and it is precisely indicated in the paper~\cite{neutrpiezo} and in
the three patents~\cite{patent1,patent2,patent3}, the parameters of
the cavitator and those of the reaction chamber are indeed finely
tuned and correlated. One for all, we point out that once the
frequency of ultrasounds was fixed and the material, shape and size
of the sonotrode was decided accordingly in order to guarantee the
resonance, the internal shape and dimensions of the vessel had to be
finely tuned so that the immersion of the sonotrode tip, its
distance from the bottom and from the lateral walls of the chamber
might generate the right pattern of ultrasonic vibrations in the
solution and hence bring about suitable cavitation conditions. This
means that there was a long experimental work in order to optimize
the cavitator and cavitation chamber before starting the actual
experimental campaign. From this perspective and considering again
the initial level of knowledge of piezonuclear reactions, it is very
important to know as many details as possible about the experimental
equipment and try to stick to them as much as possible in order to
try and obtain further positive results. In this sense, for
instance, it is not advisable to substitute the sonotrode-type
ultrasonic machine, like our cavitator, for an ultrasonic cleaner as
it was done in~\cite{canada} and hope to achieve positive results
straight away. This change implies a dramatic variation of the
cavitation conditions that do not guarantee an immediate success of
the experiments. Besides, we believe that, in an ultrasonic bath, it
is more complicated to identify the volumes and regions of the
solution where cavitation suitably takes place, due to the not easy
determination of the Chladni patterns of the bottom plate of the
basin. Moreover, so far, there is only one evidence of positive
outcomes (unfortunately not published) by an ultrasonic cleaner
which was working at frequencies higher than 0.5 MHz. Let us now
clarify one thing about the effects of cavitation on the nuclei of
the atomic species present in the solution. According to our
phenomenological model of piezonuclear reaction, the atomic species
are entrapped on the surface of the bubbles of gas in the solution
and when these bubbles are made collapse by ultrasounds, these atoms
get accelerated against each other and a piezonuclear reaction takes
place if a suitable~\cite{carmin2,carmin3} concentration of energy
in space and time is reached. From this perspective, it becomes
clear that these atoms and hence the solution involved have to be
directly subjected to ultrasounds and cavitation without any screen
between them and the ultrasonic pressure. In this sense, the
experimental equipment used by a Canadian research
team~\cite{canada} is not suitable to achieve proper cavitation
conditions to start piezonuclear reactions. Even if the used
ultrasonic cleaner at 44 KHz and 250 Watts were suitably tuned to
bring about proper cavitation conditions (energy space and time
conditions) the enclosing and sealing of the solution of Thorium in
a plastic container immersed in the ultrasonic bath would prevent
any cavitation from taking place in the container and hence from
effectively affect piezonuclearly the nuclei of Thorium. Eventually,
there is one more last remark that is useful to point out. As to
what we have said so far, it turns out that these experiments are
only apparently simple and easy to replicate. In this sense, should
one be interested in attempting to perform them, we strongly advise
not to begin with that concerning the piezonuclear decay of
Thorium~\cite{thor1}. This kind of experiment presents three
difficulties: the facing of a new experiment whose controlling
variables and parameters are still almost unknown and whose outcomes
can be not easily detectable; the presence of a radioactive
substance which certainly unpredictably affects these variables and
parameters; the safety issue due to radioactivity which makes the
replication of the attempts dangerous and more awkward. Conversely,
we strongly advise to focus the attention on those experiments whose
outcomes were the emission of neutron bursts from solutions of
stable nuclides like Iron subjected to ultrasounds and
cavitation~\cite{piezoneutr,neutrpiezo}\footnote{Even if it is
clearly reported in these references, we prefer to stress that in
these experiments, neutrons are emitted in bursts which are not
easily detectable by active neutron counters. We advise to begin
with passive integrating detectors like bubble detectors or CR39
screened by Boron. Moreover, even when one moves to electronic
detectors, it is always a sound idea to perform the measurements
both by them and passive ones, since the latter are more suitable to
detect an emission of neutrons which is certainly not constant in
time.}. As to our experience, these experiments do not present any
strong safety issue, since, apart from the emission of neutrons, we
have not noticed any activation or residual radioactivity of the
solutions and of the whole experimental equipment. If one
concentrates on these experiments and succeeds in obtaining positive
evidences of neutron emission, at the same time one accumulates
experience in governing the experimental set-up in order to produce
piezonuclear reactions which are supposed to bring about the
anomalous reduction of Thorium content. Only then does it become
advisable to move to experiments with radioactive substances.

\end{document}